\documentclass[12pt,preprint]{aastex} 

\slugcomment{{\bf ApJ, In Press}}

\shorttitle{Formation environment of the Sun}

\shortauthors{Gounelle \& Meibom}

\begin{document}

\title{The origin of short-lived radionuclides and \\
the astrophysical environment of solar system formation}

\author{Matthieu Gounelle \& Anders Meibom}
\email{gounelle@mnhn.fr}
\affil{Laboratoire d'\'{E}tude la Mati\`ere
Extraterrestre, Mus\'{e}um National d'Histoire Naturelle, \\
57 rue Cuvier, 75 005 Paris, France}

\begin{abstract}
Based on early solar system abundances of short-lived radionuclides
(SRs), such as $^{26}$Al (T$_{1/2} = 0.74$ Myr) and $^{60}$Fe
(T$_{1/2} = 1.5$ Myr), it is often asserted that the Sun was born in
a large stellar cluster, where a massive star contaminated the
protoplanetary disk with freshly nucleosynthesized isotopes from its
supernova (SN) explosion. To account for the inferred initial solar
system abundances of short-lived radionuclides, this supernova had
to be close ($\sim$ 0.3 pc) to the young ($\leqslant$ 1 Myr)
protoplanetary disk.

Here we show that massive star evolution timescales are too long,
compared to typical timescales of star formation in embedded
clusters, for them to explode as supernovae within the lifetimes of
nearby disks. This is especially true in an Orion Nebular Cluster
(ONC)-type of setting, where the most massive star will explode as a
supernova $\sim$ 5 Myr after the onset of star formation, when
nearby disks will have already suffered substantial photoevaporation
and/or formed large planetesimals.

We quantify the probability for {\it any} protoplanetary disk  to
receive SRs from a nearby supernova at the level observed in the
early solar system. Key constraints on our estimate are: (1) SRs
have to be injected into a newly formed ($\leqslant$ 1 Myr) disk,
(2) the disk has to survive UV photoevaporation, and (3) the
protoplanetary disk must be situated in an enrichment zone
permitting SR injection at the solar system level without disk
disruption. The probability of protoplanetary disk contamination by
a supernova ejecta is, in the most favorable case, 3 $\times$
10$^{-3}$.


\end{abstract}


\section{Introduction}

Short-lived radionuclides (SRs) are radioactive elements with
half-lives of the order of 1 Myr. Their presence in the nascent
solar system is inferred from an excess of their daughter isotopes
in meteorite components (Russell, Gounelle \& Hutchison 2001). Some
SRs [$^{10}$Be (T$_{1/2}$ = 1.5 Myr), $^{26}$Al (T$_{1/2}$ = 0.74
Myr), $^{36}$Cl (T$_{1/2}$ = 0.30 Myr), $^{41}$Ca (T$_{1/2}$ = 0.10
Myr), $^{53}$Mn (T$_{1/2}$ = 3.7 Myr), $^{60}$Fe (T$_{1/2}$ = 1.5
Myr)] were present in the protoplanetary disk at abundances
substantially higher than the levels expected for average
interstellar medium (Meyer \& Clayton 2000). These SRs cannot
therefore be inherited from the interstellar medium and require a
last minute origin (e.g. Wadhwa et al. 2007). The origin of SRs is
highly debated in the cosmochemistry community, because it has
important consequences for e.g. early solar system chronology (e.g.
McKeegan \& Davis 2004; Gounelle \& Russell 2005a,b) and
planetesimal heating (Urey 1955; Ghosh \& McSeen 1998). In addition,
understanding the origin of SRs will provide key constraints on the
astrophysical environment in which our solar system was born (e.g.
Hester \& Desch 2005; Goswami et al. 2005; Gounelle 2006).

Generally speaking, SRs can be made either by thermal
nucleosynthesis, at KeV energies in the interior of stars, or by
non-thermal nucleosynthesis, i.e. nuclear reactions involving cosmic
rays or accelerated particles at MeV energy from the Sun. There are
therefore two possible sources for short-lived radionuclides in the
solar protoplanetary disk: (a) Injection into the molecular cloud
core or protoplanetary disk from a nearby, late-type star (e.g.
Busso, Gallino \& Wasserburg 2003), (b) {\it In situ} production in
the protoplanetary disk by irradiation of dust and gas with
accelerated particles from the proto-Sun (e.g. Gounelle et al.
2006).

We focus here on the possible delivery of SRs by a late type star.
Asymptotic Giant Branch (AGB) stars are possible candidates
(Wasserburg et al. 2006), but this source of SRs in the early solar
system is considered highly unlikely because of the low probability
of an encounter between an AGB star and a star forming region
(Kastner \& Myers 1994). Type II supernovae represent an alternative
to AGB stars as a source of SRs (Cameron et al. 1995).

Of models based on a supernova (SN) origin for SRs there are two
types. Either a SN injects freshly synthesized SRs into a nearby
molecular cloud core, triggering its gravitational collapse (Cameron
\& Truran 1977; Cameron et al. 1995; Boss \& Vanhala 2000), or
directly into a nearby protoplanetary disk (Chevalier 2000; Hester
\& Desch 2005; Ouellette et al. 2005). The first SN scenario is now
considered less likely because only very specific conditions allow a
supernova shockwave to trigger the collapse of a molecular cloud
core and, at the same time, inject SRs (e.g. Boss \& Vanhala 2000).
In the second scenario, which is currently receiving a lot of
attention (Chevalier 2000; Hester \& Desch 2005; Ouellette et al.
2005; Ouellette, Desch \& Hester 2007), the SN has to be very close
($\sim$ 0.3 pc) to the protoplanetary disk in order to allow the
disk to intercept enough SN ejecta to account for the solar system
inventory of SRs. It is thus assumed that the massive star, which
evolved into a SN, and the protoplanetary disk were coeval and
formed in the same stellar cluster (e.g. Hester \& Desch 2005).
Massive stars form in large stellar clusters (Lada \& Lada 2003).
The nearby SN injection model therefore implies that our Sun was
born in a large stellar cluster. The Orion Nebula Cluster (ONC) is
often cited as a good analog for such an environment (Hester \&
Desch 2005).

However, if the Sun was born in a stellar cluster massive enough to
contain a SN, it has important implications for the childhood of our
solar system. Massive stars are known to power winds carving large
bubbles of hot gas in surrounding cold molecular gas (Weaver et al.
1977). They also emit large UV fluxes, which create HII blisters in
cold molecular gas (O'Dell 2001) and photoevaporate nearby
protoplanetary disks (e.g. Johnstone, Bally \& Hollenbach 1998). The
presence of numerous massive stars close to planetary systems can
also dynamically modify the orbits of the giant planets and of
Kuiper Belt Objects (e.g. Adams \& Laughlin 2001). It is therefore
essential to evaluate the {\it a priori} probability of our Sun to
be born in such an harsh astrophysical setting.

Adams \& Laughlin (2001) estimated the probability of the Sun to
have been born in a stellar environment large enough to contain a
massive star (which could deliver SRs via a SN explosion) and small
enough to preserve both the orbits of the giant planets as well as
those of the Kuiper belt objects. These authors made the assumption
that {\it any} star more massive than 25 M$_{\sun}$ was able to
deliver SRs to the solar protoplanetary disk, and they did not place
constraints on the time window during which SRs were injected into
the disk. The spatial structure of the cluster was not included in
this model. Subsequently, Williams \& Gaidos (2007) provided an
estimate for the probability of the nearby SN injection scenario.
However this estimate did not take quantitatively into account the
disk photoevaporation and overestimated the time window during which
SN injection of SRs in a nearby protoplanetary disk can occur.

The present work is organized as follows. First, we review the
cosmochemical constraints on the origin of SRs (\S 2). Second, we
present recent observations and models of molecular cloud dynamics,
star formation rate, accretion disk lifetime and evolutionary
timescales of massive stars pertinent to an evaluation of the
proposed ONC-like setting for the formation of the solar system (\S
3). Third, we calculate the probability for a protoplanetary disk
attached to a low-mass star to receive SRs from a nearby SN at the
levels inferred for the early solar system, with the necessary
constraints that SR delivery happens within the time window
specified by the cosmochemical data, and that the disk is not
destroyed by UV photoevaporation (\S 4). Our model assumptions and
limitations are discussed in section \S 5. In section \S 6 we
present the implications of our findings for the origin of
short-lived radionuclides.


\section{Polluting our protoplanetary disk with a SN ejecta: Timing
and distance} \label{sec-prolog}

Among the SRs that could have been delivered by a supernova to the
protoplanetary disk around the young Sun, $^{26}$Al provides the
strongest constraint on the timing of such an event (e.g. Wadhwa et
al. 2007; Russell et al. 2006). Aluminium-26 is more abundant in
primitive chondrite components, such as Calcium-, Aluminium-rich
Inclusions (CAIs) and chondrules (e.g. Galy et al. 2000) than in
more evolved asteroidal crustal rocks (e.g. Srinivasan et al. 1999).
CAIs have the oldest measured Pb-Pb ages of any solid matter in the
solar system (Bouvier et al. 2007). Some iron meteorites, believed
to sample the cores of differentiated asteroids, have model Hf-W
ages as old as those of CAIs (Markowski et al. 2006). They must
therefore have formed contemporaneously with CAIs. This implies that
$^{26}$Al, which is believed to have been the main heat source
responsible for asteroidal differentiation, was present in the
protoplanetary disk already at the time of iron meteorite
parent-asteroid accretion (Bottke et al. 2006). Both observations
suggest that $^{26}$Al was present in the protoplanetary disk from
the early stages of its evolution, certainly as far back in time as
the meteorite record can bring us (e.g. Wadhwa et al. 2007; Russell
et al. 2006). Although the exact timescale is not known, it is
reasonable to assume that if $^{26}$Al was delivered to the solar
system from a SN source, this injection occurred during the class 0
or class I stage of our protoSun (e.g. Montmerle et al. 2006). The
combined duration of the class 0 and class I phases is $\sim$ 10$^5$
yr. Here we will conservatively assume that the SN delivery of SRs
to the protoplanetary disk occurred within a time interval,
$\Theta$, of $\sim$ 1 Myr after the formation of the disk (see also
Ouellette et al. 2005; Looney, Tobin \& Fields 2006).

The distance $r_{}$ between the SN and the protoplanetary disk
cannot be too small, otherwise the disk will be destroyed by the
impact of the ejecta. Nor can this distance be too large because
this would limit the amount of injected SRs. There is a restricted
distance interval, the 'minimum' ($r_{\rm min}$) and the 'maximum'
($r_{\rm max}$) acceptable distance between the SN and
protoplanetary disk, referred to as the 'enrichment zone', which
satisfies these two important constraints.

Chevalier (2000) estimated that at distances smaller than $0.25$ pc,
momentum transfer between the SN ejecta and the disk will cause the
later to be stripped away. Based on 2D numerical simulations,
Ouellette, Desch \& Hester (2007) concluded that a disk as close as
0.1 pc from a 25 M$_{\sun}$ SN can survive disruption. For the
minimum distance, $r_{\rm min}$, we will adopt a conservative value
of 0.1 pc.

The maximum distance $r_{\rm max}$, below which the protoplanetary
disk receives enough SRs to account for solar system abundances
depends on a limited number of parameters. Though it is beyond the
scope of the present paper to explicitly derive $r_{\rm max}$, we
indicate how it can be calculated and summarize the estimates given
by several groups (Ouellette et al. 2005; Looney, Tobin \& Fields
2006).

For any radionuclide, the mixing fraction, $f$, of the supernova
ejecta mixed into the protoplanetary disk is:
\begin{equation}f=\frac{\rm M_{\rm SS}}{\eta \rm Y_{\rm SN}} \times
\rm e^{\Delta/\tau}, \label{eq-f}\end{equation} where $\rm M_{\rm
SS}$ is the initial mass of the radionuclide in the solar system (in
M$_{\sun}$), Y$_{\rm SN}$ is the total yield of the radionuclide
from the considered supernova (in M$_{\sun}$), $\tau $ is the mean
life, $\eta$ the injection efficiency and $\Delta$ is the time
interval between the end of nucleosynthesis in the supernova and
incorporation into refractory meteorite components (CAIs). Note that
$f \ll 1$ as $\rm M_{\rm SS} \ll \rm Y_{\rm SN}$ and $\Delta \approx
\tau$ (see below). Following Cameron et al. (1995), we translate the
mixing fraction $f$ into a geometrical relationship, $r =
\frac{r_0}{2} \sqrt{1/f}$, where $r_0$ is the size of the disk and
$r$ the distance between the disk and the massive star. Putting
these two expressions together, we obtain the general expression:
\begin{equation}r =\frac{r_0}{2} \sqrt{\eta \frac{\rm
Y_{SN}}{{\rm M}_{\rm SS}} \rm e^{-\Delta/\tau_{\rm}}}.
\label{eq-r}\end{equation}

$\rm M_{\rm SS}$ is estimated from the disk mass, the solar
abundances and the initial abundance of the considered radionuclide
identified to that of CAIs (e.g. Gounelle \& Meibom 2007). Using SN
yields calculated by state-of-the art nucleosynthetic models
corresponding to a large range of progenitor masses and
metallicities (Rauscher et al. 2002; Woosley \& Weaver 1995), and
assuming an injection efficiency of 1, Looney, Tobin \& Fields
(2006) as well as Ouellette et al. (2005) calculate that the solar
protoplanetary disk had to be at a distance between 0.02 and 0.3 pc
from the massive star for disk sizes varying between 30 and 100 AU.
$\Delta$ is constrained by the decay of short-lived radionuclides
and is of the order of $\sim$ 1 Myr. Note that the disk size cannot
be significantly larger than 100 AU because, before any injection
occurs, the disk has been sitting nearby a strongly ionizing O star
for some Myr (see section \ref{sec-Orion}), resulting in severe disk
truncation (Ouellette et al. 2005). The alternative case of larger
disks examined by Looney, Tobin \& Fields (2006) is therefore not
relevant here. For more realistic values of $\eta$ (i.e. $\eta \ll$
1) the injection distance is reduced to values approaching the
minimum acceptable distance (0.1 pc), which would cause a problem
with regard to disk survivability. We will consider here that a
generous maximum distance between the protoplanetary disk and the
supernova is $r_{\rm max}$ = 0.4 pc.



Therefore, the acceptable distance interval for which the SN
explosion delivers SRs at the level observed in the solar system,
but does not destroy the protoplanetary disk is quite narrow. The
enrichment zone lies therefore between $r_{\rm min}$ = 0.1 pc and
$r_{\rm max}$ = 0.4 pc.

\section{The unlikely ONC-like setting}
\label{sec-Orion}

The Orion Nebular Cluster is the epitome of a massive star forming
region in the solar neighborhood, often invoked as an analog for the
birthplace of our Sun (Hester \& Desch 2005; Ouellette et al. 2005;
Bally, Moeckel \& Throop 2005). The ONC is dominated by the four
bright Trapezium O and B stars, which are named $\theta ^1$ and
numbered from A to D (O'Dell 2001). It is situated $\sim$ 450 pc
away from the Sun and contains roughly half a dozen stars massive
enough to end their lives as SN (Hillenbrand 1997). In the ONC, the
most massive object is the O7 star $\theta ^1${\it C} Ori, which has
a mass of $\sim 40$ M$_{\sun}$. Within $1$ pc of $\theta ^1${\it C}
Ori are thousands of low-mass protostars, $70$\% of which have disks
(Hillenbrand et al. 1998). Some of the disks are as close as a few
tenths of a parsec from $\theta ^1${\it C} Ori.

\begin{figure}
\epsscale{0.8} \plotone{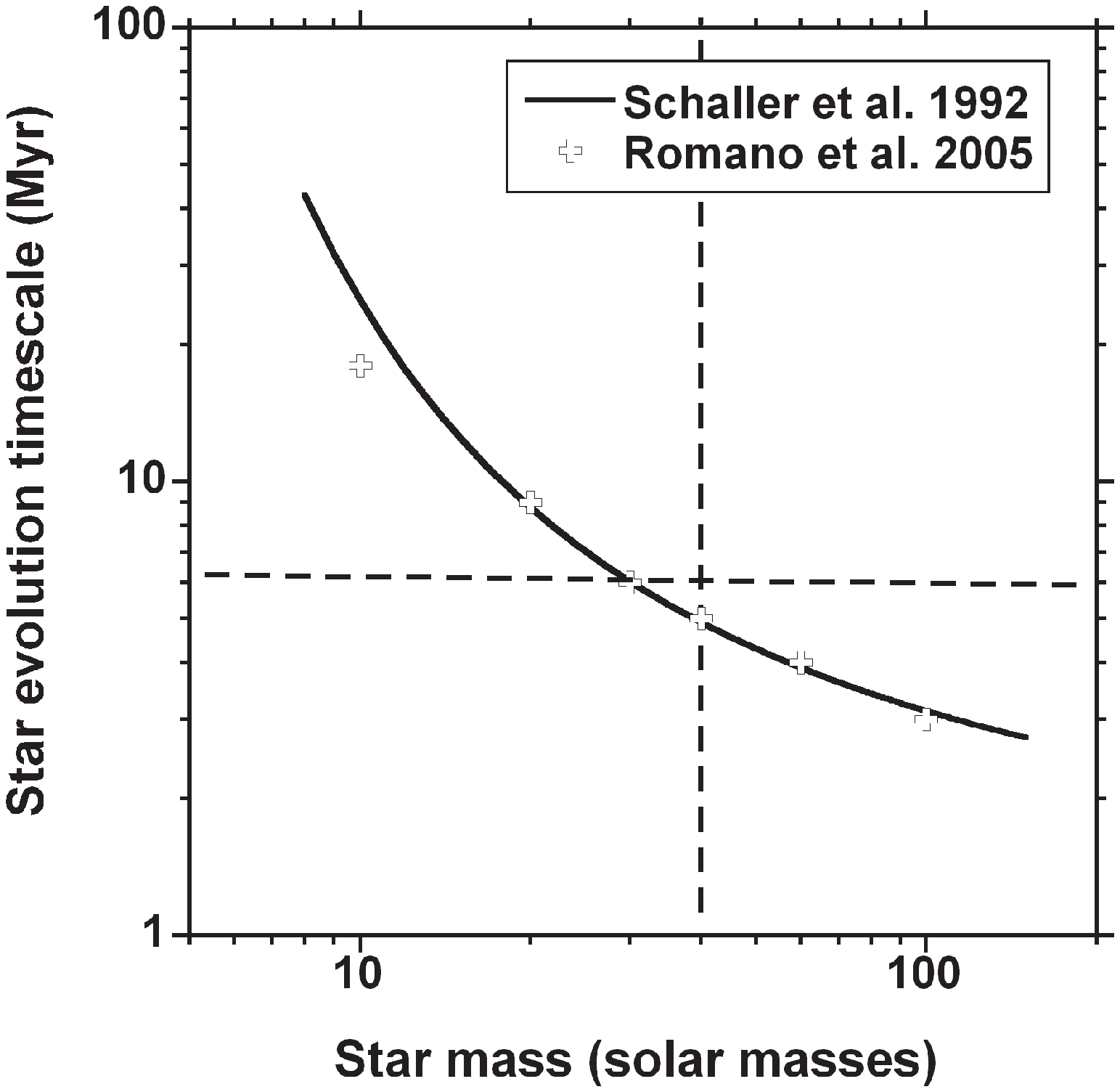} \caption{Evolutionary timescales of
massive stars. Schaller et al. (1992) fitted their models with the
expression ${\rm log}({\rm t}_{\rm SN})={1.4}/{({\rm log } M_{\rm
SN})^{1.5}}$ where $\rm t_{\rm SN}$ and $ M_{\rm SN}$ are
respectively the evolutionary timescale (in Myr) and the mass of the
considered star (in M$_{\sun}$). More recent work by Romano et al.
(2005) confirm the findings of Schaller et al. (1992).}
\label{fig-schaller}
\end{figure}

Though evolving on substantially shorter timescales than low-mass
stars, high mass stars still need a finite time to complete hydrogen
and helium burning, and explode in a SN (Schaller et al. 1992;
Romano et al. 2005). It takes $\sim 40$ Myr for a $8$ M$_{\sun}$ B3
star and $\sim 3$ Myr for an extremely massive $120$ M$_{\sun}$ O3
star to reach the SN stage respectively. A $40$ M$_{\sun}$ star,
such as $\theta ^1${\it C} Ori, evolves for $\sim 5$ Myr before it
goes supernova (Figure \ref{fig-schaller}).

The age of the ONC is estimated to be $\leqslant$ 1 Myr (e.g.
Hillenbrand 1997). The age of $\theta ^1${\it C} Ori is not
precisely known, although this star is probably among the youngest
members of the ONC, because early formation of $\theta ^1${\it C}
Ori would have halted star formation through rapid photoionization
of the surrounding gas (O'Dell 2001; Boss \& Goswami 2007). O'Dell
(2001) argues that $\theta ^1${\it C} Ori is {\it the} youngest star
in the ONC cluster. Palla \& Stalher (2001) determined the age of
another Trapezium star, $\theta ^1${\it D} Ori, and find an upper
limit of 0.1 Myr. Here it is conservatively assumed that $\theta
^1${\it C} Ori is 1 Myr old and will go SN 5 Myr after the onset of
star formation in the ONC, i.e. $\sim$ 4 Myr from now.

The protoplanetary disks that are currently within a few tenths of a
parsec from $\theta ^1${\it C} Ori will suffer severe mass-loss due
to photoevaporation driven by the UV radiation from this and other
massive stars in the region (Johnstone, Hollenbach \& Bally 1998;
St\"{o}rzer \& Hollenbach 1999; Richling \& Yoke 2000; Balog et al.
2007). Typical mass loss rates due to photoevaporation are 10$^{-7}$
M$_{\sun}$/yr (Johnstone, Hollenbach \& Bally 1998; St\"{o}rzer \&
Hollenbach 1999). The typical lifetime of a minimum mass solar
nebula of mass 0.01 M$_{\sun}$ (Hayashi et al. 1985) in such an
environment is therefore $\sim$ 10$^5$ yr. Johnstone, Hollenbach \&
Bally (1998) studied the specific case of $\theta ^1${\it C} Ori and
showed that, at a distance of 0.3 pc, a protoplanetary disk would
shrink to a size of about 1 AU in about 1 Myr, i.e. well before
$\theta ^1${\it C} Ori becomes a SN. Disk evaporation simply
prevents injection of SN ejecta because disks in the vicinity of the
SN have essentially disappeared before the SN explosion takes place.

It is possible that disks adjacent to $\theta ^1${\it C} Ori could
partially survive photo-evaporation if their orbits around the
massive star are highly elliptical rather than circular (e.g. Adams
\& Bloch 2005), or if the photo-evaporation rate decreases rapidly
with time (e.g. Adams et al. 2006). However, in 4 Myr from now, when
$\theta ^1${\it C} Ori goes supernova, such surviving disks will be
highly evolved, harboring large planetesimals as well as giant
planets (e.g. Montmerle et al. 2006). In an Orion-like setting,
planetary formation is accelerated by photoevaporation (Throop \&
Bally 2005), reducing accordingly the time window for SR injection.
Disks which survive evaporation will therefore not receive SRs from
the SN explosion until very late in their evolution, inconsistent
with cosmochemical evidence for early delivery, which indicates that
the injection happened within 1 Myr of disk formation (see section
\ref{sec-prolog}).

Over the next $4$ million years, there will be few new low-mass
stars forming within 1 pc of $\theta ^1${\it C} Ori. Star formation
is certainly more vigorous during the first Myr of a molecular cloud
lifetime and decreases sharply with time (see Lada \& Lada 2003).
After a few Myr, the gas in a molecular cloud has dissipated, and
new star formation no longer takes place (Leisawitz, Bash \&
Thaddeus 1989; Elmegreen 2000; Hartmann, Ballesteros-Paredes \&
Bergin 2001; Lada \& Lada 2003). Winds from massive stars and
outflows from low-mass stars are primarily responsible for the quick
dissipation of the molecular gas (Bally, Moeckel \& Throop 2005). In
fact, young clusters, such as NGC1333 and Serpens, contain
relatively large populations of Class 0 sources and are rich in
bipolar flows, both features indicative of the earliest stages of
star formation (e.g. Knee \& Sandell 2000; Sandell \& Knee 2001). In
contrast, older clusters, such as IC348, contain few protostellar
sources and outflows (Lada \& Lada 2003). The fraction of disks in a
given cluster, which is homologous to the star formation rate, also
steeply decreases with the cluster age (see Figure 1 of Haisch, Lada
\& Lada 2001).

The decrease with time of the star forming rate is even more
dramatic in the immediate vicinity of a massive star ($\lesssim$ few
pc) than at the global molecular cloud scale. Hillenbrand (1997)
argues that star formation in the Trapezium, i.e. in the vicinity of
$\theta ^1${\it C} Ori, is now over. More generally, star formation
cannot occur in the absence of molecular gas, which is ionized by
the enhanced UV flux of massive stars. Wren \& O'Dell (1995)
estimate that the ionized HII region created by $\theta ^1${\it C}
Ori is now $0.3$ pc wide and is growing at the rate of $0.5$ km/s.
Within the next $4$ Myr, this region will expand to a $\sim$ $2.3$
pc wide HII cavity around $\theta ^1${\it C} Ori, where star
formation will be effectively halted. Hillenbrand \& Hartmann (1998)
estimate that the $\sim$ 1 pc wide gaseous region surrounding
$\theta ^1${\it C} Ori will be photoevaporated on a timescale of
$\leqslant$ 1 Myr.

It is not only the ionizing power of massive stars, such as $\theta
^1${\it C} Ori, that contributes to the sharp decrease in star
formation rate in their immediate vicinity. The winds emanating from
the massive stars create bubbles. The linear dimension of a wind
bubble increases with the mass of the star (Chevalier 1999). A 20
M$_{\sun}$ star creates a cavity as large as 11 pc in 7 Myr
(Chevalier 1999), suggesting that the molecular gas surrounding
$\theta ^1${\it C} Ori will be dissipated by winds in addition to
photoionization on timescales significantly shorter than the
lifetime of $\theta ^1${\it C} Ori.

After a few Myr of evolution, star formation in ONC-like settings
occurs mainly in photo-dissociation regions (PDRs) at the interface
of the HII region and the molecular gas (Healy, Hester \& Claussen
2004), a few parsecs away from the massive star. The fate of the ONC
is well illustrated by the 2-3 Myr old NGC2244 cluster, whose most
massive star, HD206267, has the same spectral type as $\theta
^1${\it C} Ori. In NGC2244, star formation is occurring in the
outskirts of the cluster, at distances 5-10 pc from the central O6
star (Hartmann 2005). In addition, Reach et al. (2004) note that
these stars, which formed $\sim$ 2 Myr before HD206267 goes
supernova, are the last generation of stars forming in the cluster.
Low-mass stars formed 3 Myr after the onset of star formation in a
given molecular cloud are too far away to be contaminated by SRs at
any significant level.

\begin{figure}
\epsscale{.80} \plotone{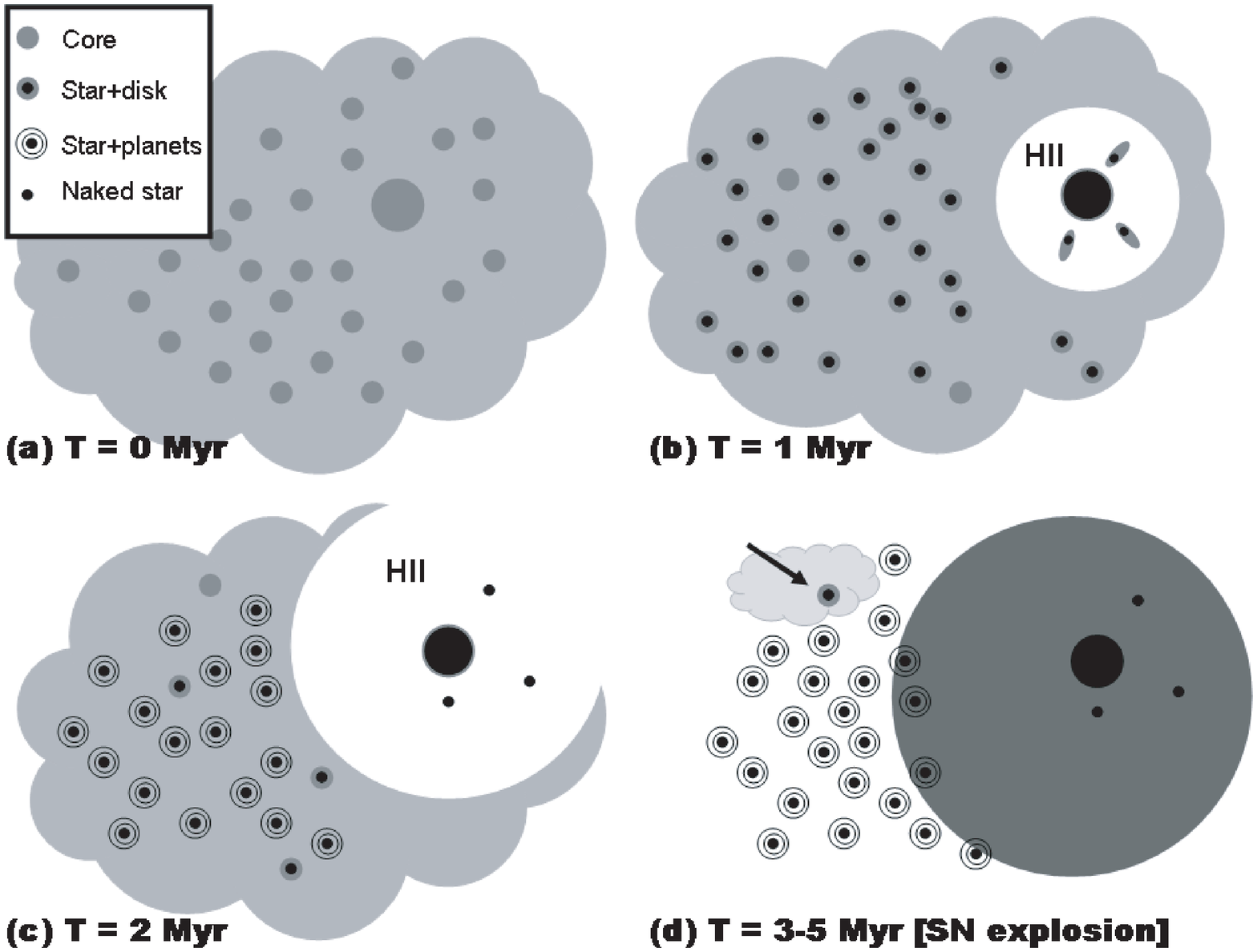} \caption{Sketch illustrating the
fundamental problems met by the idea of short-lived radionuclides
injection in a protoplanetary disk by a nearby supernova. In this
cartoon, the evolution of a massive star (large circle) and several
low-mass stars within molecular gas (light grey) are shown. In panel
(a) is depicted the onset of star formation assumed to be coeval for
all stars. During that stage, only dense cores are present. Note
that if there is any spread in the onset of star formation, it would
correspond to a delayed formation of the massive star (see text).
One million year later (panel (b)), the massive star has carved a
$\sim$ 0.3 pc wide HII cavity. Disks which are within this cavity
are evaporating as depicted by their cometary appearance similar to
that observed in Orion protoplanetary disks. Star formation is less
vigorous than at T = 0 Myr. At T = 2 Myr (panel (c)), the HII cavity
has expanded to $\sim$ 1 pc (see text). The disks which are inside
the cavity are totally photoevaporated (Johnstone, Hollenbach \&
Bally 1998). Low-mass stars that formed at T = 0 Myr now have
planets. Very few new stars are forming (see text and Figure
\ref{fig-psi}). When the massive star goes supernova (panel (d)),
the stars in the uttermost vicinity of the supernova do {\it not}
have disks, while disks further away have already dissipated to form
planets. At this stage, star formation is virtually non-existent
because molecular gas has largely been cleaned by the massive star
wind and UV emission. A left-over, evanescent, parcel of molecular
gas is shown in light grey. Injection of SRs by the supernova in a
late-formed disk (marked with an arrow) is an extremely improbable
event. An indicative scale would be given by the size (few tenths of
parsec) of the HII cavity in panel (b).} \label{fig-sketch}
\end{figure}

Figure \ref{fig-sketch} illustrates the discrepancy between the
timescales of massive star evolution and low-mass star formation for
an ONC-like setting.



\section{Probability of SR injection in a young disk by a nearby SN}
In this section, we calculate two numbers: (1) Within a cluster of a
given size $N_*$, what is the probability for a young disk
($\leqslant$ 1 Myr) attached to a low-mass star to be contaminated
in SRs by a nearby supernova explosion at the level observed in the
early solar system? (2) What is the probability for {\it any}
protoplanetary disk in the Galaxy to be contaminated in SRs from a
nearby SN explosion at the level observed in the solar system? The
latter number will be calculated by integrating the former with the
cluster size distribution (section \ref{sec-any}).


The probability for a disk surrounding a low-mass star in a cluster
of a given size ($N_*$) to be enriched in SRs by a nearby supernova
at the observed solar system abundance without being disrupted by UV
photoevaporation is given by the expression:
\begin{equation} P_{\rm SR} (N_*)
=\int^{M_u}_{M_l} f_{\rm YSO} ({\rm t}_{\rm SN})~f_{\rm enrichment}~
f_{\rm survivors}~ {\cal P}_{M_{\rm SN}}(N_*) dM_{\rm SN},
\label{eq-P1}
\end{equation} where $f_{\rm YSO}$ is the fraction of young
($\leqslant$ 1 Myr) low mass stars (or Young Stellar Objects)
present in the cluster at the time ($\rm t_{\rm SN}$) of the
supernova explosion, $f_{\rm enrichment}$ is the fraction of
low-mass stars present in the enrichment zone, $f_{\rm survivors}$
is the fraction of disks which have survived UV photoevaporation and
${\cal P}_{M_{\rm SN}}(N_*)$ is the probability that the most
massive star in a cluster containing $N_*$ stars has a mass $M_{\rm
SN}$. M$_l$ = 8 $M_{\sun}$  and $M_u$ = 150 M$_{\sun}$ are,
respectively, the masses of the least and most massive star possibly
going supernova (Kroupa \& Weidner 2005). The dependence of $f_{\rm
YSOs}$, $f_{\rm enrichment}$, $f_{\rm survivors}$ and ${\cal
P}_{M_{\rm SN}}$ on $N_*$ and $M_{\rm SN}$ will be described in the
following sections.


\subsection{The most massive star in a cluster}
\label{sec-proba} 
\begin{figure}
\epsscale{0.8} \plotone{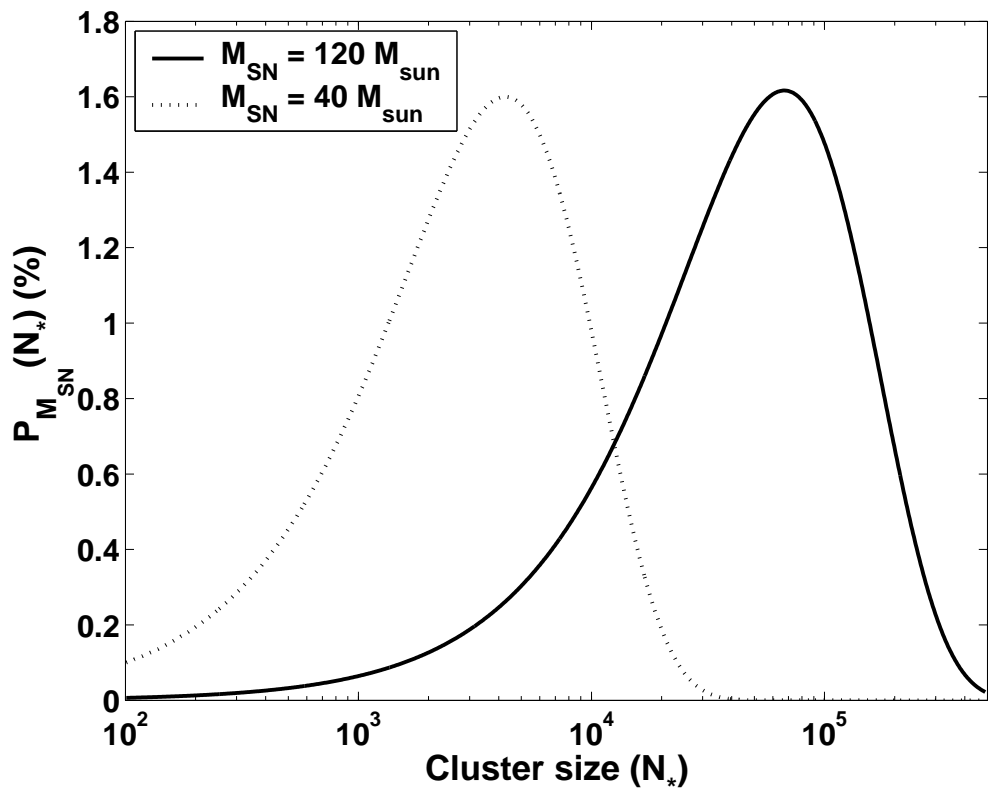} \caption{Probability that the most
massive star in a cluster containing $N_*$ stars has a mass $M_{\rm
SN}$ = 40 M$_{\sun}$ and $M_{\rm SN}$ = 120 M$_{\sun}$ (equation
(\ref{eq-calP})). Note that the maximum probability for a 40
M$_{\sun}$ star to be the most massive star occurs for a cluster of
$\sim$ 4000 stars, slightly larger than the ONC.} \label{fig-calP}
\end{figure}

Using the Initial Mass Function (IMF), dN/dM
$\propto$ M$^{-(1+\alpha)}$, where N is the number of stars and M
the star mass and with $\alpha$ = 1.5 (Scalo 1986), Williams \&
Gaidos (2007) show that the probability that the most massive star
in a cluster containing $N_*$ stars has a mass $M_{\rm SN}$ is:
\begin{equation} {\cal P}_{M_{\rm SN}}(N_*) =\frac{\alpha f_{\rm SN}
N_*}{[(M_u/M_l)^{\alpha}-1]M_u}\left(\frac{M_u}{M_{\rm
SN}}\right)^{1+\alpha}e^{-{\cal N}( >M_{\rm SN})}, \label{eq-calP}
\end{equation}
where $f_{\rm SN}$ = 0.003 is the total fraction of stars going
supernova, i.e. having a mass larger than 8 M$_{\sun}$ (Adams \&
Laughlin 2001). $\cal N(> \it M_{\rm SN})$, the number of stars with
mass larger than $M_{\rm SN}$, is obtained from the IMF and given by
the expression:
\begin{equation}
{\cal N} ( > M_{\rm SN})= f_{\rm SN} N_* \frac{(M_u/M_{\rm
SN})^{\alpha}-1}{(M_u/M_l)^{\alpha}-1}. \label{eq-massive}
\end{equation}
The dependence of ${\cal P}_{M_{\rm SN}}(N_*)$ on N$_*$ is shown in
Figure \ref{fig-calP} for two fiducial masses, $M_{\rm SN}$ = 40
M$_{\sun}$ and $M_{\rm SN}$ = 120 M$_{\sun}$.

\subsection{Fraction of stars in the enrichment zone}
\label{sec-fe}
 Young stellar clusters are segregated in mass, with
the more massive stars present at the center of the cluster
(Hillenbrand \& Hartmann 1998 and references therein). This
core-halo structure appears to be primordial, based on morphological
(Larson 1982) and dynamical (Hillenbrand \& Hartmann 1998) evidence.
It is thus reasonable to consider that clusters have a spherical
symmetry, and that the massive star responsible for the putative
injection of SRs is placed at the center of the cluster (see for
example Rho et al. 2006). The number of stars in the enrichment zone
is therefore given by the expression:
\begin{equation}f_{\rm enrichment} (N_*)={\int_{r_{\rm
min}}^{r_{\rm max}} \rho(r) 4 \pi r^2 dr}  / \\
{\int_{0}^{{\rm R}_{c} (N_*)} \rho(r) 4 \pi r^2 dr}, \label{eq-fe}
\end{equation}
where $\rho (r)$ is the stellar volume density and $r_{\rm min}$ and
$r_{\rm max}$ define respectively the minimum and maximum acceptable
distance between the disk and the SN, defining the enrichment zone
(see section \ref{sec-prolog}). $N_*$ and R$_c$($N_*$) are the
number of stars contained in the cluster and the cluster radius,
respectively.

Observations of ONC as well as IC348 show that the numbers of stars
per unit area, i.e. the internal cluster density distribution,
decreases approximately as $1/r$ (Hillenbrand \& Hartmann 1998;
Muench et al. 2003), in which case the stellar volume density, $\rho
(r)$, goes as 1/$r^{2}$. Inserting the latter expression into
equation ($\ref{eq-fe}$) yields:
\begin{equation}
f_{\rm enrichment} (N_*)=\frac{r_{\rm max}-r_{\rm min}}{{\rm
R}_c(N_*)}.\label{eq-fe2}
\end{equation}

Using the embedded clusters compilation of Lada \& Lada (2003),
Adams et al. (2006) find that embedded clusters follow the law ${\rm
R}_c \approx {\rm R}_{300} \sqrt{N_*/300}$, where R$_c$ is the
cluster size and R$_{300}$ is an empirical parameter varying between
1 and 3 pc (Figure 2 of Adams et al. 2006). This provides an
estimate of the average surface density of clusters of different
sizes, which can be formulated as:
\begin{equation}
\Sigma = {N_*}/{\pi {\rm r}_c^2}. \label{eq-sig}
\end{equation}
We consider a wide range for $\Sigma$ encompassing the range
observed by Adams et al. (2006), i.e. $\Sigma$ varies from $10$ to
$300$ stars/pc$^2$. Note that the ONC is characterized by
$\Sigma_{\rm ONC}$ = 211 stars/pc$^2$, based on the cluster
parameters (R$_c$ = 2.06 pc and N$_*$ = 2817 stars) of Hillenbrand
\& Hartmann (1998). Inserting equation (\ref{eq-sig}) into equation
(\ref{eq-fe2}) yields:
\begin{equation}
f_{\rm enrichment} (N_*) = \frac{(r_{\rm max}-r_{\rm min})\sqrt{\pi
\Sigma} }{\sqrt{N_*}}.\label{eq-fe3}
\end{equation}


\subsection{Fraction of disks surviving photoevaporation}


Disks residing within a few tenths of a parsec of a massive O star
will be UV photoevaporated on timescales significantly shorter than
1 Myr (e.g. St{\"o}rzer \& Hollenbach 1999). Because disks can have
elliptical orbits around the massive star, and because
disk-shrinking is limited by the decrease of the mass loss rate with
the disk size (Adams et al. 2004), a non-zero fraction of disks
might survive nearby massive stars. To estimate the fraction of
disks that survive total photoevaporation we will turn again to
Orion.

Using the Hubble Space Telescope planetary camera, Johnstone,
Hollenbach \& Bally (1998) measured the size of 28 protoplanetary
disks in the vicinity of $\theta ^1${\it C} Ori. Only nine of them
have sizes larger than 35 AU, the estimated minimum size of the
solar protoplanetary disk (e.g. Gomes et al. 2005; Hartmann 2005).
This means that, in the case of the $\sim$ 1 Myr old ONC, only a
disk fraction $f_{\rm survivors}$ = 0.32 has resisted to
photoevaporation and are eligible to SN pollution. This number is
compatible with the depletion of disks by a factor of $\sim$ 3
observed by Balog et al. (2007) within 0.5 pc of O stars in the 2-3
Myr NGC 2244 star forming region.


\subsection{Fraction of young low-mass stars at time t$_{\rm SN}$}
\label{sec-YSO}
 As discussed in section \ref{sec-prolog},
cosmochemical constraints require the SN injection to occur early in
the evolution of the protoplanetary disk. We conservatively
estimated that injection had to occur in the first 1 Myr of disk
evolution. The fraction of low-mass stars younger than $\Theta $ = 1
Myr at time t$_{\rm }$ is:
\begin{equation}
\left\{
\begin{array}{lllr}
 f_{\rm YSO} (\rm t) &=& \int_{0}^{\rm t_{\rm
}}\psi(t')dt'~ /~\int_0^{\infty}\psi(t')dt' & \rm for~ t\leqslant
\Theta \\
 \vspace{10 pt}
f_{\rm YSO} (\rm t) &=& \int_{\rm t_{\rm }-\Theta}^{\rm t_{\rm
}}\psi(t')dt'~ /~\int_0^{\infty}\psi(t')dt' & \rm for~ t\geqslant
\Theta
\end{array}
\right. \label{eq-fd}
\end{equation}
where $\psi$(t) is the star forming rate.

Star forming rates are not very well constrained and are
the subject of debate among astronomers. We therefore consider three
different expressions for $\psi$(t), a combination that is likely to
bracket reality.

The first expression for $\psi$(t) is a step function with a total
star formation timescale T = 4 Myr: \begin{equation} \left\{
\begin{array}{lllr}
 \psi (t)& =& \psi_1 ~ &\rm for~ t\leqslant T \\
 \vspace{5pt}
\psi(t) & = &0 ~ & \rm for ~t \geqslant T
\end{array}
\right. \label{eq-psi1}
\end{equation}
Expression (\ref{eq-psi1}) gives an overestimated measure of the
star formation rate, because star forming regions are known to be
more active during their first million years than during subsequent
times (e.g. Hartmann, Ballesteros-Paredes \& Bergin 2001).

The second expression for $\psi$(t) is a linearly decreasing
function, terminating on a timescale of T = $4$ Myr, due to
molecular gas dissipation (Leisawitz, Bash \& Thaddeus 1989;
Elmegreen 2000; Hartmann, Ballesteros-Paredes \& Bergin 2001; Lada
\& Lada 2003).
\begin{equation}
\left\{
\begin{array}{lllr}
 \psi (t) &= &\psi_2 \times (1-\frac{t}{T}) ~ &{\rm for} ~
t\leqslant T \\
\vspace{5pt}
 \psi(t) &= &0 ~ &{\rm for} ~ t \geqslant T
\end{array}
\right.
\end{equation}

\begin{figure}
\epsscale{0.8} \plotone{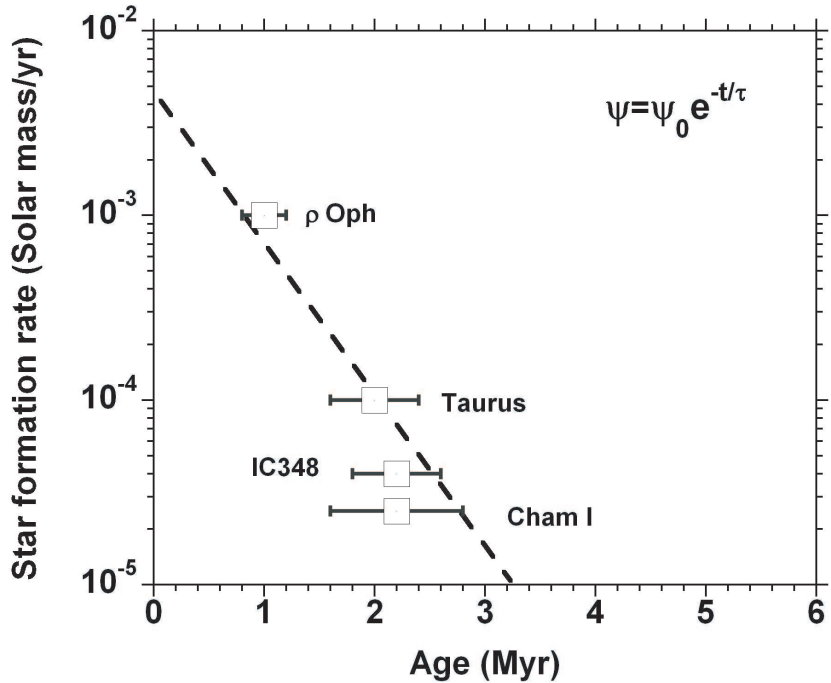} \caption{Star-formation rate versus
age for a suite of well studied molecular clouds. Cham and $\rho$
Oph stand for Chamaeleon I and $\rho$ Ophiuchi respectively. The
thick dashed line is the fit of the data which minimizes the
exponential decrease of the star formation rate (see section
\ref{sec-YSO}). It gives a decay time constant of $\tau$ = 0.56 Myr.
The age of a given star-forming region depends on the mass of the
stars and their position on the Hertzsprung-Russell diagram, among
other parameters (Palla \& Stahler 2000; Hartmann 2001). For $\rho$
Ophiuchi, we take the commonly accepted age of $\sim$ 1 Myr
(Preibich \& Zinnecker 1999; Palla \& Stahler 2000). For other
molecular clouds, we adopt the age compilation as well as their
associated error bars given by Haisch, Lada \& Lada (2001). These
ages are compatible with other, more recent, estimates in the
literature (e.g. Luhman 2004; Muench et al. 2007). The errors take
into account the error in the mean of the source age derived from a
given set of Pre Main Sequence (PMS) tracks (Haisch, Lada \& Lada
2001). Errors due to the use of different PMS models are of the
order of $1$ Myr and could result in a shift of the data without
changing the relative age sequence (Haisch, Lada \& Lada 2001). Star
formation rates for individual molecular clouds are given by
Preibisch \& Zinnecker (1999) without associated error bars. Note
that star formation rates are not available for older clusters
because there are no observed star-forming molecular clouds older
than a few Myr (Elmegreen 2000; Lada \& Lada 2003; Hartmann,
Ballesteros-Paredes \& Bergin 2001).} \label{fig-psi}
\end{figure}

 The third expression for $\psi$(t) is an exponentially
decreasing function with a characteristic star formation timescale
of $\tau$ = 0.56 Myr:
\begin{equation}
\psi (t) = \psi_0 \times e^{-\frac{t}{\tau}} \label{eq-psi3}
\end{equation}
Expression (\ref{eq-psi3}) is determined from a fit of star
formation rates vs. ages for a suite of observed clusters shown in
Figure \ref{fig-psi}. We emphasize that all the star forming regions
considered in this plot have similar sizes (i.e. similar contents of
gas). This expression describes a rather extreme, though possible,
star formation history with most stars being formed during the first
million year of evolution.

Our preferred model corresponds to the linearly decreasing law for
$\psi(t)$ as it describes an intermediary situation between the two
extreme cases, the step function and the exponentially decreasing
law.

\begin{figure}
\epsscale{0.8} \plotone{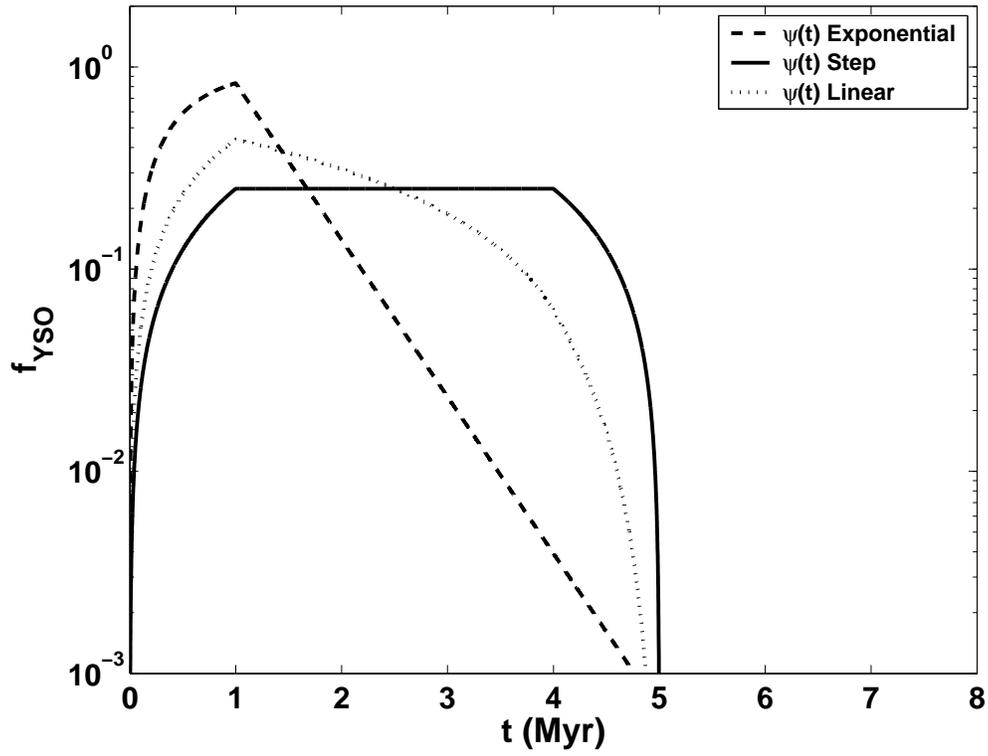} \caption{Fraction of low-mass stars
younger than 1 Myr ($f_{\rm YSO}$) present in the cluster as a
function of time t (equation \ref{eq-fd}). $f_{\rm YSO}$ is shown
for the 3 expressions considered for the star forming rate,
$\psi$(t), and discussed in the text.} \label{fig-fd}
\end{figure}

The fraction of low-mass stars (calculated with equation
(\ref{eq-fd})) present in the cluster as a function of time, $f_{\rm
YSO} (\rm t_{\rm })$, is shown in Figure \ref{fig-fd} for the three
different star formation rates considered. The sharp cut-offs at
$\sim$ 5 Myr in Figure \ref{fig-fd} for the linear and the step
function star formation rates are due to finite duration of star
formation in an embedded cluster (Lada \& Lada 2003 and references
therein). For these cases, there are no low-mass stars young enough
to be enriched in the cluster at a time t = T + $\Theta$ = 5 Myr.

For disk contamination to occur, the relevant quantity is the
fraction of low-mass stars at the time t$_{\rm SN}$ of the supernova
explosion, i.e. $f_{\rm YSO}(\rm t_{\rm SN})$. Evolutionary models
for massive stars (Schaller et al. 1992; Romano et al. 2005)
provide\ the time t$_{\rm SN}$ at which a massive star goes
supernova through the expression:
\begin{equation}{\rm log}({\rm t}_{\rm SN})={1.4}/{({\rm log }M_{\rm SN})^{1.5}},
\label{eq-time}\end{equation} where $\rm t_{\rm SN}$ and $M_{\rm
SN}$ are expressed in Myr and in solar masses respectively.

\subsection{Injection probability in a cluster containing $N_*$ stars: Results}

\begin{figure}
\epsscale{0.8} \plotone{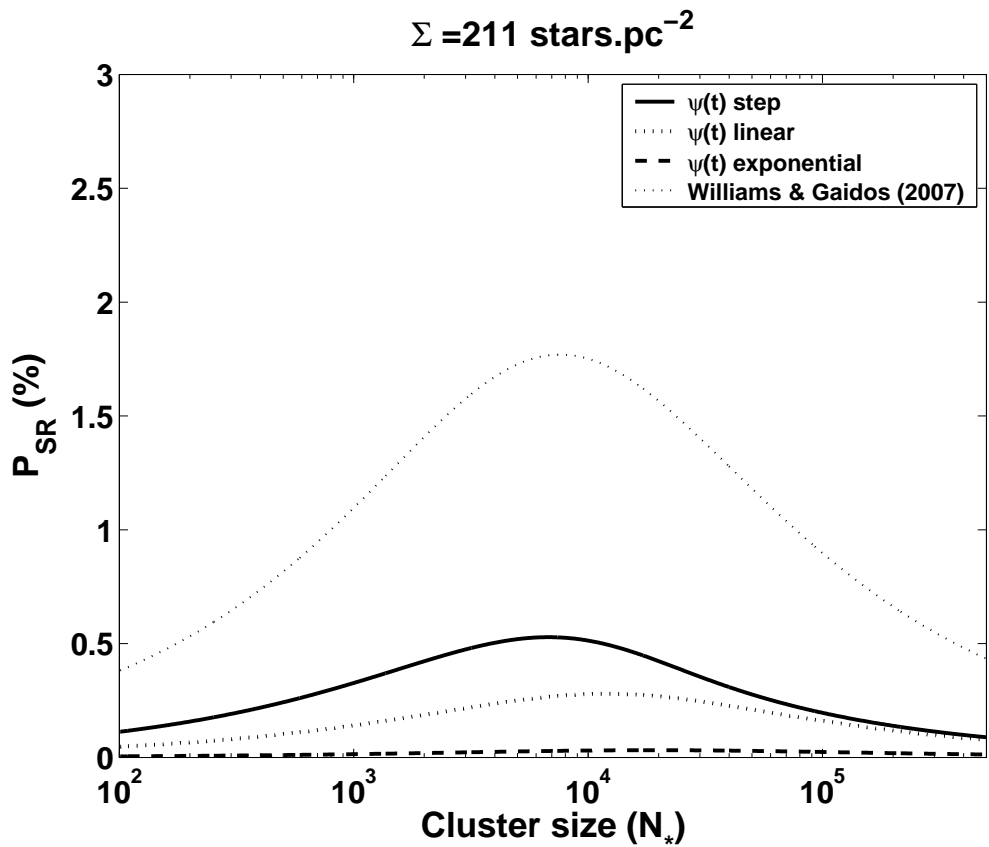} \caption{Probability of
protoplanetary disk enrichment in SRs at the level observed in the
solar system ($P_{\rm SR} (N_*)$) as a function of the cluster size
($N_*$), and calculated for the different expressions of the star
forming rate $\psi$(t). The curves were obtained using the ONC
stellar surface density of $\Sigma_{\rm ONC}$ = 211 stars/pc$^2$. A
similar curve obtained by Williams \& Gaidos (2007) is shown for
comparison (with $\Sigma _3$ = 211 stars/pc$^2$).}
\label{fig-results}
\end{figure}

Putting together the expressions defined above, we obtain:
\begin{equation}
P_{\rm SR} (N_*) ={\it f}_{\rm survivors}\int^{M_u}_{M_l} f_{\rm
YSO}({\rm t}_{\rm SN})~\frac{(r_{\rm max}-r_{\rm min})\sqrt{\pi
\Sigma} }{\sqrt{N_*}} ~  {\cal P}_{{\it M}_{\rm SN}}(N_*) d{\it
M}_{\rm SN}, \label{eq-sum}
\end{equation}
where $f_{\rm YSO}$(t$_{\rm SN}$) and ${\cal P}_{M_{\rm SN}}(N_*)$
are given by equations (\ref{eq-fd}), (\ref{eq-time}) and
(\ref{eq-calP}) respectively. With $\Sigma = \Sigma_{\rm ONC}$ = 211
stars/pc$^2$ (section \ref{sec-fe}) and $r_{\rm min}$ = 0.1 pc and
$r_{\rm max}$ = 0.4 pc (section \ref{sec-prolog}), we obtain a set
of curves for $P_{\rm SR} (N_*)$ shown in Figure \ref{fig-results}.
Our curves are slightly below that calculated by Williams \& Gaidos
(2007) because we take explicitly into account the disk
photoevaporation, and because we model the timing of injection in a
fashion dictated by the cosmochemical constraints, i.e. we consider
that injection happened during the first Myr of disk evolution
rather than at any time (section 2).

The probability for SR enrichment peaks at $\sim$ 10 000 stars. It
is however very low as it reaches a maximum value of $\sim$ 0.4 \%
for the most favorable case, corresponding to a step function law
for the star formation rate $\psi$(t). It peaks at $\sim$ 0.03 \%
for the exponential form of the star formation rate, which is based
on the data shown in Figure \ref{fig-psi}. Our preferred model (a
linearly decreasing function for $\psi$(t)) gives an intermediary
result with a maximum value of $\sim$ 0.3 \%. Increasing $\Sigma$ to
its maximum value, $\Sigma$ = 300 stars/pc$^2$, does not
substantially change these results, because  $P_{\rm SR} (N_*)$
varies smoothly with the square root of $\Sigma$ (see equation
(\ref{eq-sum})) and all our calculations are normalized to the {\it
high} stellar density of the ONC.


\subsection{Probability of injection for {\it any} disk}

\label{sec-any} 
\begin{figure}
\epsscale{0.8} \plotone{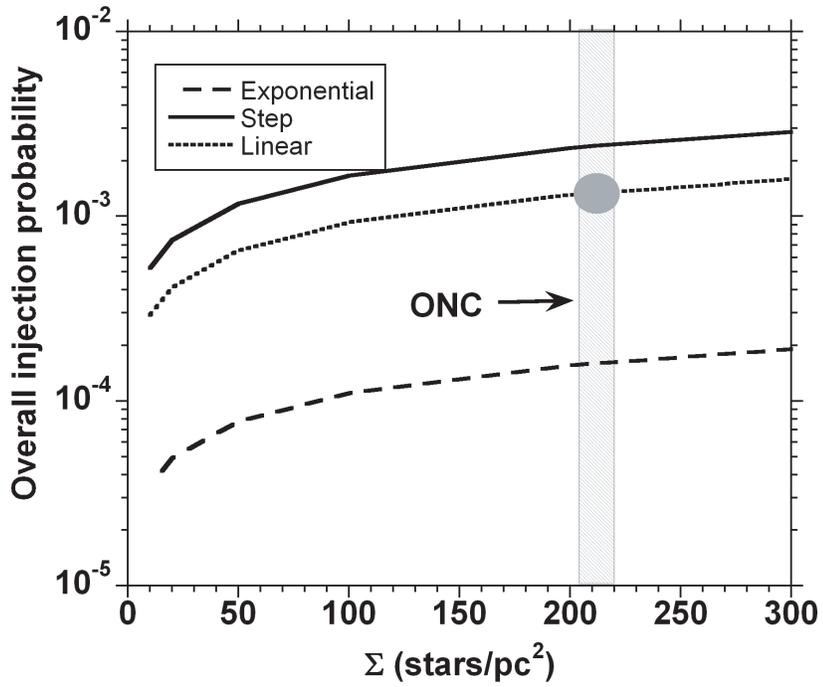} \caption{Probability for {\it any}
protoplanetary disk attached to a young low-mass star to receive SRs
from a nearby SN at the level observed in the solar system. The
vertical line denotes the surface density of the ONC, while the grey
circle represents our preferred model, with the Orion surface
density and a linear decrease of $\psi$(t).} \label{fig-gal}
\end{figure}

To calculate the probability of SR injection for
{\it any} protoplanetary disk attached to a low-mass star in the
Galaxy, ${\cal P}_{\rm GAL}$, we will integrate $P_{\rm SR} (N_*)$
over the probability for a star to form in a cluster of size $N_*$,
which scales as 1/$N_*$ (Elmegreen \& Efremov 1997; Adams \&
Laughlin 2001).  In other words,
\begin{equation}
{\cal P}_{\rm GAL}={\cal K}\int^{\rm N_{\rm max}}_{\rm N_{\rm min}}
P_{\rm SR} (N_*) \frac{dN_*}{N_*}, \label{eq-gal} \end{equation}
where N$_{\rm min}$ = 100 and N$_{\rm max}$ = 5 $\times$ 10$^5$ are
the minimal and maximal sizes of stellar clusters respectively
(McKee \& Williams 1997; Williams \& Gaidos 2007). The normalization
constant, ${\cal K} = 9.39 \times 10^{-2}$, is calculated using the
observation that $\sim$ 80 \% of stars form in clusters larger than
100 members (Lada \& Lada 2003). Figure \ref{fig-gal} shows ${\cal
P}_{\rm GAL}$ for the three different star forming rates discussed
in section \ref{sec-YSO}. In our preferred case, where $\Sigma =
\Sigma_{\rm ONC}$ = 211 stars/pc$^{2}$ and $\psi$(t) follows a
linearly decreasing law, the probability for any disk to receive SRs
from a nearby SN at the level observed in the solar system is 1
$\times$ 10$^{-3}$ (grey circle in Figure \ref{fig-gal}). This
probability increases with the square root of $\Sigma$ as expected
from equation (\ref{eq-sum}). In the most favorable case, i.e.
adopting a step function for the star formation rate and with
$\Sigma$ = 300 stars/pc$^{2}$, the probability for any diskS to
receive SRs from a nearby SN at the level observed in the solar
system is 3 $\times$ 10$^{-3}$.

\subsection{Multiple supernovae}
\label{sec-msn}
\begin{figure}
\epsscale{0.8} \plotone{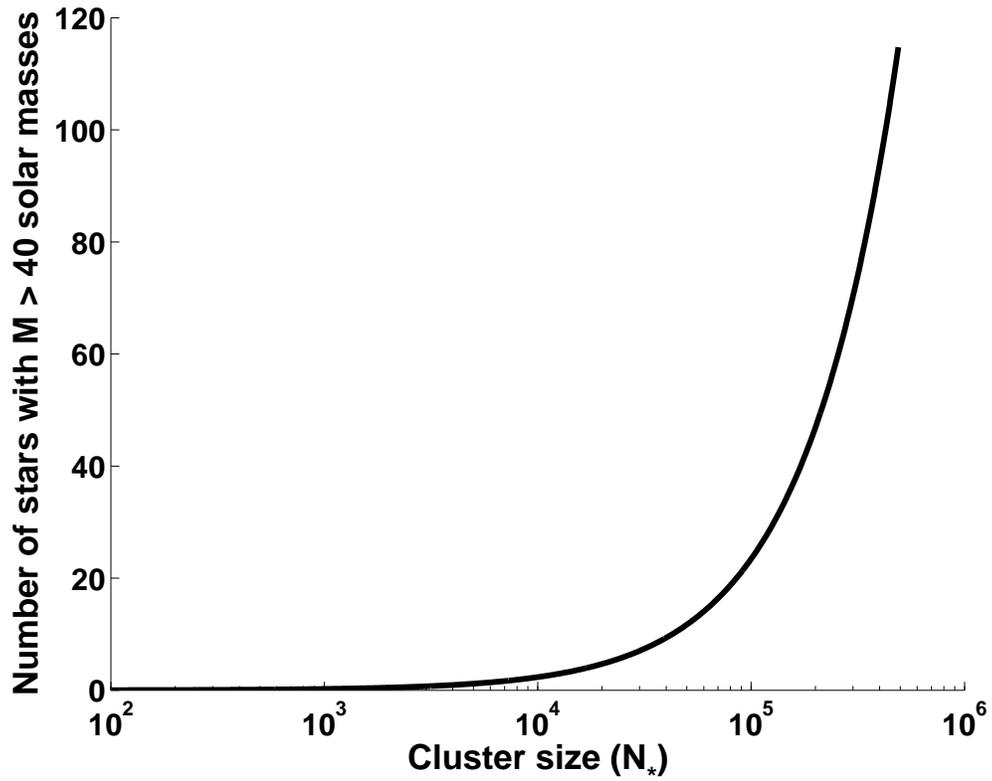} \caption{Number of stars with mass
larger than 40 M$_{\sun}$ as a function of cluster size (calculated
from equation (\ref{eq-massive})). Multiple supernovae ($\geqslant$
2) capable to inject SRs in a cluster occur only in clusters
containing more than 10$^4$ stars.} \label{fig-msn}
\end{figure}

 If the cluster size is large enough, multiple
supernovae can occur within the lifetime of the cluster. Stars with
mass below 40 M$_{\sun}$ can not contribute to the inventory of SRs
because they need too long timescales to evolve (see section
\ref{sec-Orion}). Multiple supernovae injection of SRs therefore
occur only in clusters large enough to contain at least one star
whose mass is larger than 40 M$_{\sun}$. This condition is satisfied
only for clusters containing more than $N_*$ = 10$^4$ stars
(Equation (\ref{eq-massive}) and Figure \ref{fig-msn}).

To calculate the {\it maximum} contribution of nearby massive stars
to a young protoplanetary disk, we assume that, for clusters
containing more than $N_*$ = 10$^4$ stars, every low-mass star will
witness a supernova explosion in its youth ($\leqslant$ 1 Myr), i.e.
that:
\begin{equation} \int^{M_u}_{M_l} f_{\rm
YSO} ({\rm t}_{\rm SN}) {\cal P}_{M_{\rm SN}}(N_*) dM_{\rm SN} =1.
\label{eq-msn}
\end{equation}
Introducing this condition in equation (\ref{eq-sum}) yields a
stringent upper limit, as this statement is probably true only for
the most massive clusters (N$_*$ $\geqslant$ 10$^5$), where there
are enough stars to include tens of supernovae whose progenitors
stars have masses larger than 40 M$_{\sun}$. With this generous
permission, we calculate the probability for any young low-mass star
protoplanetary disk to be contaminated in SRs by a nearby supernova
explosion to be 6 $\times$ 10$^{-3}$ in the most favorable case
($\psi$(t) described by a step function and $\Sigma$ = 300
stars/pc$^2$).


\section{Discussion}

\subsection{Beyond Orion: The Carina Nebula}
\label{sec-car}

The probability of a protoplanetary disk to receive SRs from a
nearby SN explosion would be higher if one considered a star forming
region more massive than the ONC. Such a region would contain more
massive stars than the ONC according to the initial mass function
(Salpeter 1955). For example, a $120$ M$_{\sun}$ star would be ready
to go SN and deliver SRs to a nearby disk after only $\sim$ $3$ Myr
of evolution (Figure \ref{fig-schaller} and equation
(\ref{eq-time})). But, $3$ Myr after the onset of star formation,
the star formation rate has already decreased significantly as can
be appreciated in Figure \ref{fig-psi}. Additionally, extremely
massive molecular clouds are rare, at least in the vicinity of the
Sun. Within $2$ kpc of the Sun, where the number of embedded
clusters is large enough to make statistically robust observations,
the ONC is the most massive example (Lada \& Lada 2003). Finally, a
more massive cluster contains more massive stars which are more
efficient in dissipating molecular gas by ionization and
photo-dissociation (e.g. Bally, Moeckel \& Throop 2005). Despite
these difficulties, the massive Carina Nebula star-forming region is
now proposed by some authors (e.g. Smith \& Brooks 2007), instead of
Orion, as a good analog for the astrophysical environment of our
solar system formation.

The Carina Nebula (NGC3372) is a star forming region situated at 2.3
kpc from the Sun and containing $\sim$ 65 O stars (Smith 2006).
$\eta$ Carinae, the most massive star contained in the Carina Nebula
($\geqslant$ 100 M$_{\sun}$, Davidson \& Humphreys 1997), needs
$\sim$~3 Myr to go supernova (Smith 2006). Since the age of the
cluster is $\sim$ 3 Myr, assuming it formed at the beginning of the
cluster's life, $\eta$ Carinae can go supernova anytime now (Smith
2006). Most of the molecular gas in the vicinity of $\eta$ Carinae
has however been cleared away, creating an extended HII region where
star formation is halted (Smith et al. 2000 and Figure 5 of de
Graauw et al. 1981).

A few evaporating protoplanetary disks were tentatively detected in
the Carina Nebula and await confirmation (Smith, Bally \& Morse
2003). From the coordinates given by Smith, Bally \& Morse (2003),
we calculate that the protoplanetary disk closest to $\eta$ Carinae
lies at a minimum distance of 2.4 pc, a factor of $\sim$ 10 too
large compared to the distance needed to inject SRs in sufficient
number. In addition, the putative protoplanetary disks detected by
Smith, Bally \& Morse (2003) are significantly more massive (by a
factor of $\sim$ 100) than the protoplanetary disks observed in
Orion (Johnstone, Hollenbach \& Bally 1998), or than the expected
solar protoplanetary disk (0.013 M$_{\sun}$, Hayashi et al. 1985),
suggesting that they belong to young intermediate-mass stars rather
than to solar-like stars.

The scarcity of solar-like protoplanetary disks in the vicinity of
$\eta$ Carinae is due to the extreme ionizing power of that star
which emits $Q(H) \sim 10^{50.6}$ ionizing photons per second, which
can be compared to the $Q(H) \sim 10^{49}$ photons per second
emitted by $\theta ^1${\it C} Ori (O'Dell 2001). It confirms that in
the immediate vicinity of very massive stars, protoplanetary disks
are extremely rare because (1) those accompanying low-mass stars
which formed contemporaneously with massive stars were
photoevaporated or formed planets (Throop \& Bally 2005) and (2) the
efficient gas clearing by winds and UV emission has halted new
low-mass star formation.

Triggered star formation in the Carina Nebula occurs in the gas-rich
region named South Pillars (Smith \& Brooks 2007). Smith \& Brooks
(2007) argue that young stars in the South Pillars region will be
pelted by supernova ejecta containing SRs. While it is true that
these young stars will be exposed to the supernova ejecta, the South
Pillars lie 20 pc away from $\eta$ Carinae (Smith et al. 2000),
almost two orders of magnitude further away than the distance
required to incorporate SRs at the abundance observed in the solar
system (see section \ref{sec-prolog}). Massive stars which form now
in the South Pillars will need several Myr to explode and deliver
SRs to the interstellar medium and therefore suffer from the same
difficulties discussed above.


\subsection{On our probability estimate}
\label{sec-dis2}

In calculating the different probability estimates for the injection
of SRs into a young protoplanetary disk attached to a low-mass star,
we made several assumptions and simplifications that we discuss in
the following.

\subsubsection{The enrichment zone}
We used a minimum distance between the supernova and the disk,
$r_{\rm min}= 0.1 $ pc, proposed by Ouellette, Desch \& Hester
(2007) from models simulating the interaction of a 25 M$_{\sun}$ SN
with a protoplanetary disk.
A more massive star, more likely to achieve that goal, would have a
more massive ejecta and therefore a more disruptive effect (through
momentum transfer) on the disk (Chevalier 2000). The minimum SN-disk
distance we adopted is therefore a lower limit, which turns into an
upper limit for $f_{\rm enrichment}$ (see equation \ref{eq-P1}) and
therefore into an upper limit for the probability of SR delivery by
a nearby SN in a young disk.

The distance $r$ varies with the square root of the injection
efficiency ($\eta$) and with the inverse square root of the initial
abundance of short-lived radionuclides ($\rm M_{SS}$) (equation
(\ref{eq-r})). The initial abundance of SRs is known within a factor
of a few (Gounelle 2006). Therefore, changes in their adopted
abundance will not change much the results. In the case of injection
into a protoplanetary disk, Ouellette, Desch \& Hester (2007) show
that the injection efficiency is below 1 \% for the gas fraction of
the ejecta. Though they suggest that SRs could be injected in the
disk in the form of dust grains, it remains to be demonstrated how
plausible dust injection is, especially given that SN dust
condensation efficiency (defined as the ratio of mass of refractory
elements condensed into dust to that of refractory elements in the
ejecta) is $\leqslant$ 0.12 (Sugerman et al. 2006 and references
therein). Ercolano et al. (2007) estimate an upper limit as low as 4
$\times$ 10$^{-3}$ for the dust condensation efficiency of supernova
SN 1987A. If $\eta$ is reduced to one order of magnitude in equation
(\ref{eq-r}), it will reduce the maximum acceptable distance by a
factor of $\sqrt 10 \sim 3$, i.e. from 0.4 pc to $\sim$ 0.15 pc. The
maximum distance $r_{\rm max}$ calculated in section
\ref{sec-prolog} is thereore a strict upper limit, which turns into
an upper limit for $f_{\rm enrichment}$ (see equation \ref{eq-P1})
and, accordingly, for the probability of SR delivery by a nearby SN
in a young disk.

\subsubsection{Disk UV photoevaporation}
To estimate the fraction of disks which survived UV
photoevaporation($f_{\rm survivors}$), we calculated from the data
of Johnstone, Bally \& Hollenbach (1998) the fraction of disks in
Orion which are larger than the minimum inferred original size of
the solar protoplanetary disk (35 AU, Gomes et al. 2005). This
provides a strong upper limit for $f_{\rm survivors}$ for three
reasons. First, the ONC is only $\sim$ 1 Myr old and at the time of
$\theta ^1${\it C} Ori explosion (4 Myr from now) many more disks
will have their sizes reduced below 35 AU. Second, this estimate
does not include the disks which have already fully evaporated and
will not capture any SRs. Third, more massive stars, more likely to
deliver SRs, have also larger UV fluxes which will result in a more
rapid photoevaporation (Johnstone, Bally \& Hollenbach 1998). The
fraction of disks that survive UV photoevaporation, $f_{\rm
survivors}$, which we used in our calculations is therefore an upper
limit, as is accordingly the probability of SR delivery by a nearby
SN in a young disk.

\subsubsection{Temporal coincidence of disks and supernovae in a
molecular cloud}


In calculating $f_{\rm YSO}$, we considered $\psi$(t), the temporal
evolution of the star formation rate, at the global molecular cloud
scale, and did {\it not} take into account the dissipation of
molecular gas in the immediate vicinity of the most massive star.
This is a simplification as discussed in section \ref{sec-Orion}.
Because massive stars create very rapidly after their formation a
HII region in which star formation is impossible, star formation in
molecular clouds occurs in the outskirts of HII regions, at several
parsecs from the massive star (Reach et al. 2004; Balog et al. 2007;
Hartmann 2005). In other words, after 2-3 Myr of evolution, the star
formation rate, $\psi$(t), in the immediate vicinity ($\lesssim$ 2
pc) of any massive star is virtually zero. By assuming non-zero star
formation rates, we definitely calculated a strong upper limit for
$f_{\rm YSO}$ and therefore for the probability of contamination of
a young protoplanetary disk by a nearby supernova.

In calculating $ P_{\rm SR}(N_*)$ (equation (\ref{eq-P1})) and
$f_{\rm YSO}$ (equations (\ref{eq-fd}) and (\ref{eq-time})), we
implicitly assumed that star formation was coeval in the considered
star forming region. In fact, it is only if the most massive stars
in a cluster formed in advance relative to low-mass stars that the
probability of injection can increase. In such a case, the rapid
evolution timescale of protoplanetary disks could be reconciled with
the slower evolution timescale of massive stars (Figure
\ref{fig-sketch}) and the destructive effect of massive stars would
play a less important role.

It is commonly argued that molecular clouds have short lives ($\sim$
4 Myr) and that star formation within them proceed as soon as the
cloud is assembled (e.g. Ballesteros-Paredes et al. 2007). This
implies that the star formation age spread is small (though non
zero, e.g. Hartmann 2001). On the other hand Palla \& Stalher (2000)
have argued that molecular clouds live some tens of Myr and that
some stars form in advance compared to the majority of stars. Even
if this is correct, it would not solve the timescale problem,
because all stars which possibly formed early relative to the
majority of stars in the cluster have (at least in the case of the
ONC) masses below 0.3 M$_{\sun}$ (Palla et al. 2007). Such stars are
far from being massive enough to go SN and deliver SRs to a nearby
protoplanetary disk. We note that, in general, high-mass stars are
likely to form last in a cluster (Bally et al. 1998; St{\"o}rzer \&
Hollenbach 1999; O'Dell 2001; Kroupa \& Weidner 2005; Henriksen
1986; Boss \& Goswami 2007). This implies again that the probability
estimates given in the previous sections are upper limits.

\subsubsection{Multiple supernovae}

The contamination probability in the case of multiple supernovae was
explicitly overestimated in assuming (equation (\ref{eq-msn})) that
every low-mass star in the cluster will witness a supernova
explosion in its youth (T $\leqslant$ 1 Myr). The overall
contamination probability was also implicitly overestimated by
keeping $f_{\rm enrichment}$ identical to that calculated in the
single supernova case. This is because, in calculating $f_{\rm
enrichment}$ (section \ref{sec-fe}), we assumed that the most
massive star is located at the center of the cluster, where the
density of stars is highest. In the case of multiple supernovae, the
additional massive stars are located in the cluster periphery (as
only the most massive star occupies the cluster center) where
low-mass stars are rare, decreasing accordingly $f_{\rm
enrichment}$.

Even with those gross approximations, the effect of multiple
supernovae is to increase the probability by a factor of only a few.
This factor of a few would cancel if the physical effect of many
more massive stars were properly taken into account. More massive
stars would lead to increased photoevaporation power, clearing very
large HII regions where star formation is halted. For example the
HII region surrounding $\eta$ Carinae has an expansion radius of 110
pc (Smith et al. 2000). In addition, the presence of many more
massive stars would also increase the probability of close stellar
encounters, and therefore planetary disruption (see below). For all
these reasons, we will ignore the increase by a factor of a few in
the calculated probability due to multiple supernovae.

\subsubsection{Planetary disruption}
A low-mass star born in a large cluster where many massive stars are
present can endure close stellar encounters, which have the
possibility to disrupt the orbits of planets and small bodies. This
was modeled by Adams \& Laughlin (2001). They showed that for
clusters larger than $\sim$ 2500 members, the cluster density is
such that stellar encounters would modify the orbits of the giant
planets and the Kuiper Belt. This led Adams \& Laughlin (2001) to
suggest that the Sun formed in a cluster smaller than a few thousand
stars. In doing so, they considered relatively long relaxation
timescales for clusters. This was criticized by Hester \& Desch
(2005). Though small clusters live only up to 10 Myr before
disruption, there is a non zero fraction of clusters, corresponding
to the most massive ones, which live hundreds of Myr (Lada \& Lada
2003), similarly to the lifetimes of the Pleiades. Kroupa et al.
(2001) suggested on the basis of a dynamical study that the ONC
might become an open cluster similar to the Pleiades, reinforcing
the idea that for large clusters the relaxation timescales adopted
by Adams \& Laughlin (2001) is correct. Because other complications
might arise (e.g. Megeath et al. 2007) we did not, however, take
explicitly into account the effect of close stellar encounters in
our calculations. Neglecting these effects for large clusters
implies that the probability of SR delivery by a nearby SN to a
young disk, which we calculated above, is an upper limit.

\section{Implications for the origin of short-lived radionuclides}

\subsection{The origin of short-lived radionuclides}

The probability for any protoplanetary disk attached to a young
low-mass star to receive SRs from a nearby SN at the level observed
in the solar system is $\sim$ 1 $\times$ 10$^{-3}$ in our preferred
case. As shown in section \ref{sec-dis2}, this value is a stringent
upper limit. Although improbable does not mean impossible, this very
low probability suggests that other sources should be considered for
explaining the overabundance (compared to the interstellar medium
average value) of SRs in the early solar system (Wadhwa et al.
2007). Some of the SRs, such as $^{26}$Al, $^{36}$Cl, $^{41}$Ca and
$^{53}$Mn can be made by {\it in situ} irradiation together with
$^{7}$Be and $^{10}$Be (e.g. Gounelle et al. 2006), and therefore do
not necessitate further elucidation. Iron-60, on the other hand
cannot be made by local irradiation (Lee et al. 1998). It is
therefore necessary to find a source of $^{60}$Fe, within a
plausible astrophysical context. Besides the nearby supernova
scenario, which seems to be unlikely, there are two remaining
possibilities for the $^{60}$Fe presence in the solar system: (1) a
distant SN or (2) an inherited origin resulting from the
contributions of many SN.

A distant supernova (i.e. at a few parsecs from the solar system)
was invoked in the past as the source of $^{60}$Fe and other
short-lived radionuclides (Cameron \& Truran 1977; Cameron et al.
1995). In that context, it was also assumed that the supernova
shockwave triggered the collapse of the dense core. This proposition
has a number of problems. First, very special conditions are needed
to inject short-lived radionuclides in a dense molecular cloud core
without disrupting it (e.g. Boss \& Vanhala 2000 and references
therein). The supernova shockwave needs to impact the core with a
fine-tuned velocity of $\sim$ 20 km/s. A higher shockwave velocity
would disrupt the dense core, while a slower one would fail to
induce collapse and inject radioactivities. Second, injection
calculations do not take into account the filamentary nature of
molecular clouds, which consist of the fractal juxtaposition of high
density clumps and low density interclumps matter (e.g. Rho et al.
2006), but treat the cloud as an homogeneous and dense matter. A
supernova shockwave within a molecular cloud will follow the path of
less resistance, and will probably avoid the denser regions and
instead expand into the interclump matter (Chevalier 1999)
delivering very few, if any at all, short-lived radionuclides to
molecular cloud cores. Third, though the larger size of a molecular
cloud core compared to a protoplanetary disk allows it to collect
SRs from a SN ejecta at larger distances (see equation \ref{eq-r}),
it cannot be much further away than a few pc (Looney, Tobin \&
Fields 2006), and many of the difficulties described above hold in
that case too.

\subsection{The solar system initial content of $^{60}$Fe}

For some years, a high inferred initial $^{60}$Fe/$^{56}$Fe ratio
(e.g. Mostefaoui et al. 2005) was considered as a smoking gun for
ruling out an inherited origin for $^{60}$Fe (Hester \& Desch 2005).
The experimental situation has now changed as was recently discussed
at the Hawaii meeting on the chronology of
meteorites\footnote{http://www.lpi.usra.edu/meetings/metchron2007}.
The maximum solar system $^{60}$Fe/$^{56}$Fe initial ratio of 1
$\times$ 10$^{-6}$ (e.g. Wadwha et al. 2007) is probably too high an
estimate. Below we give a summary of the various estimates for the
initial $^{60}$Fe/$^{56}$Fe ratio in the solar system.

%

Measurements of Ni isotopes in CAIs are challenging because CAIs are
rich in Ni nucleosynthetic anomalies, which can blur the effect of
$^{60}$Fe decay (Birck 2004). An upper limit on the
$^{60}$Fe/$^{56}$Fe ratio of 1.6 $\times$ 10$^{-6}$ was given by
Birck \& Lugmair (1988) for an Allende CAI with a Fe/Ni ratio of
$\sim$ 20. We note that no isochron was reported for this object,
and that the authors expressed caution about the $^{60}$Fe decay
origin of the small ($\sim$ 0.01 \%) anomaly detected. A two point
internal isochron for an Allende CAI measured by Quitt\'{e} et al.
(2007) gave a lower limit for the initial $^{60}$Fe/$^{56}$Fe ratio
of 3 $\times$ 10$^{-7}$.

Measurements on the silicate portion of chondrules from primitive
meteorites, including Semarkona, yield $^{60}$Fe/$^{56}$Fe ratios
between 1.7 and 3.2 $\times$ 10$^{-7}$ (Tachibana et al. 2006;
Tachibana et al. 2007), while a measurement from a Semarkona
troilite (FeS) gives $^{60}$Fe/$^{56}$Fe $\sim$ 9 $\times$ 10$^{-7}$
(Mostefaoui et al. 2005). High $^{60}$Fe/$^{56}$Fe initial ratios in
troilite can be due to Fe-Ni redistribution in the sulfides during
later alteration processes (Guan et al. 2007). It is indeed
difficult to understand how troilite would have formed with a higher
initial $^{60}$Fe/$^{56}$Fe ratio than silicates in chondrules from
the same meteorite. If an hypothetic time delay of 1.6 Myr is
assumed between the formation of the solar system and the closure of
the Fe-Ni system in ordinary chondrites (Connelly et al. 2007),
initial $^{60}$Fe/$^{56}$Fe ratios between 3.6 and 6.7 $\times$
10$^{-7}$ can be calculated from the measurements of Tachibana et
al. (2006) and Tachibana et al. (2007).

Differentiated meteorites add to the complexity of the Fe-Ni system.
The absence of $^{60}$Fe evidence in minerals from eucrites and
angrites with high Fe/Ni ratios lead Sugiura et al. (2006) to put an
upper limit of 1.5 $\times$ 10$^{-7}$ on the initial
$^{60}$Fe/$^{56}$Fe ratio. Quitt\'{e} et al. (2006) reported values
of (1.3 $\pm$ 0.8) $\times$ 10$^{-9}$ and (8.2 $\pm$ 2.6) $\times$
10$^{-9}$ for the $^{60}$Fe/$^{56}$Fe at the time of formation of
the angrites Sahara-99555 and d'Orbigny. Combined with Pb-Pb ages of
$\sim$ 3.3 and $\sim$ 2.6 Myr after CAIs respectively (Amelin et al.
2006; Amelin 2007), these estimates lead to an upper value for the
initial $^{60}$Fe/$^{56}$Fe of 1 $\times$ 10$^{-7}$. The study of
the eucrites Chervony Kut and Juvinas lead Shukolyukov \& Lugmair
(1993) to put an upper limit of 2 $\times$ 10$^{-7}$ for the initial
$^{60}$Fe/$^{56}$Fe ratio.

Although there is at present no consensus on the initial value of
$^{60}$Fe in the early solar system, a low value for the
$^{60}$Fe/$^{56}$Fe ratio of a few times 10$^{-7}$ seems preferred.



\subsection{An inherited origin for $^{60}$Fe}

While a high (1 $\times$ 10$^{-6}$) initial $^{60}$Fe/$^{56}$Fe
ratio could be interpreted as a strong evidence for a nearby SN
contamination, revision to a lower initial value
($^{60}$Fe/$^{56}$Fe $\sim$ 3 $\times$ 10$^{-7}$) changes the
outcome because it is compatible with a galactic background origin
for $^{60}$Fe.

The present average galactic value of the $^{26}$Al/$^{27}$Al ratio
has been estimated to be 8.4 $\times$ 10$^{-6}$ (Diehl et al. 2006).
Recent measurements of $\gamma$-ray lines in the interstellar medium
with the Integral satellite have yielded a ratio $^{60}$Fe/$^{26}$Al
= 0.148 (Wang et al. 2007). Using an elemental ratio
$^{27}$Al/$^{56}$Fe = 0.109 (Lodders 2003) gives an average galactic
value for the $^{60}$Fe/$^{56}$Fe ratio of 1.4 $\times$ 10$^{-7}$.
This latter value is identical to the lower estimates of the early
solar system inferred from meteorites (see above). Even if the
initial solar system $^{60}$Fe/$^{56}$Fe ratio was as high as 4
$\times$ 10$^{-7}$ (e.g. Quitt\'{e} et al. 2007), fluctuations
around the average value could easily explain the presence of
$^{60}$Fe due to an inherited origin in the solar system.

It is important to note that this estimate is that of the {\it
present} interstellar medium. It might have been significantly
different 4.6 Gyr ago. In addition, this number is averaged on the
entire inner Galaxy. Heterogeneities in the $^{26}$Al abundance in
the Galaxy (Kn{\"o}dlseder et al. 1999) suggest that both $^{26}$Al
and $^{60}$Fe abundances can fluctuate around the average value
calculated above. Such calculations show that if the initial solar
system $^{60}$Fe/$^{56}$Fe ratio was only a few times 10$^{-7}$, the
need for a single supernova vanishes and an inherited origin can
account for the $^{60}$Fe found in the early solar system. Though
$^{26}$Al decays faster than $^{60}$Fe, an inherited origin cannot
be excluded for this radionuclide either\footnote{The recent claim
for a late injection of $^{60}$Fe (Bizzarro et al. 2007) was based
on $^{60}$Ni deficits (relative to chondrites) in iron meteorites
and pallasites, which failed to be confirmed by other groups
(Andrews et al. 2007; Regelous et al. 2007; Dauphas et al. 2008).
There is at present no evidence for late injection of iron-60.}.

\acknowledgments

We are especially grateful to Fred Adams and Thierry Montmerle for
their generous and stimulating inputs throughout the writing of that
paper. We thank Edouard Audit, J\'{e}r\^{o}me Bouvier, Roland Diehl,
Jean Duprat, Eric Feigelson, Lee Hartmann, Patrick Hennebelle,
Shu-Ichiro Inutsuka, Doug Johnstone, Mordecai Mac Low, Kevin
McKeegan, Alessandro Morbidelli, Manuel Petitat, Ghylaine
Quitt\'{e}, Fran\c{c}ois Robert, Sienny Shang, Frank Shu, Vincent
Tatischeff and Edward Young for illuminating discussions and sharp
criticisms. An anonymous reviewer helped to improve the manuscript.
This study made use of the ADS abstract service and was partially
funded by the Programme National de Plan\'{e}tologie (PNP) and the
CNRS fund France-Etats-Unis.

\clearpage

\clearpage

\end{document}